\documentclass[aps,prb,reprint,groupedaddress]{revtex4-1}
\usepackage{graphicx}
\usepackage{dcolumn}
\usepackage{bm}
\draft 

\begin{document}


\title{Significant suppression of thermal conductivity in FeSb$_2$ by Te doping} 



\author{Kefeng Wang}
\altaffiliation{Permanent address: Nanjing National Laboratory of Mircrostructure and Department of Physics, Nanjing University, Nanjing 210093 China}
\author{Rongwei Hu}
\altaffiliation{Present address: Ames Laboratory US DOE and Department of Physics and Astronomy, Iowa State University, Ames, IA 50011, USA}
\author{C. Petrovic}
\affiliation{Condensed Matter Physics and Materials Science Department, Brookhaven National Laboratory, Upton, New York 11973}


\date{\today}

\begin{abstract}
Kondo insulator like material FeSb$_2$ was found to exhibit colossal Seebeck coefficient. It would have had huge potential in thermoelectric applications in cryogenic temperature range if it had not been for the large thermal conductivity. Here we studied the influence of Te doping at Sb site on thermal conductivity and thermoelectric effect in high quality single crystals.
Surprisingly, only $5\%$ Te doping suppresses thermal conductivity by two orders of magnitude, which may be attributed to the substitution disorder. Te doping also results in transition from an semiconductor to a metal. Consequently  thermoelectric figure of merit $(ZT\sim0.05)$ in Fe(Sb$_{0.9}$Te$_{0.1}$)$_2$ at $\sim 100K$ was enhanced by about one order of magnitude when compared to $ZT<0.005$ in undoped FeSb$_2$.
\end{abstract}


\maketitle 

The revival of research on the solid state thermoelectric (TE) cooling and electrical power generation devices could be mainly attributed to some attractive features, such as long life, the absence of moving parts and emissions of toxic gases, low maintenance and high reliability.\cite{TE1,TE2} Present thermoelectric materials have relatively low energy conversion efficiency that can be evaluated by thermoelectric figure of merit $ZT=(S^2/\rho\kappa)\cdot T$, where $S$ is the Seebeck coefficient, $\rho$ is the electrical resistivity, $\kappa$ is the thermal conductivity and $T$ is the absolute temperature. Recent efforts to design materials with enhanced thermoelectric properties were mainly focused on reducing the lattice contribution to thermal conductivity by alloy scattering, superstructures or nanostructure engineering, \cite{phonon1, phonon2, phonon3} On the other hand, interest in the potential merits of electronic correlation effects was revived by the discovery of large Seebeck coefficients in transition metal compounds, such as FeSi~\cite{FeSi} and Na$_x$CoO$_2$~\cite{NaCoO}. In a Kondo insulator, localized \textit{f} or \textit{d} states hybridize with conduction electron states leading to the formation of a small hybridization gap with large density of states (DOS) just below and above the gape. This resonance in DOS centered about 2-3 $k_BT$ from the Fermi energy could induce very large Seebeck coefficient, as observed in FeSi. \cite{correlated1,correlated2,correlated3} Up to now, most of the effort has been concentrated on high temperature range and none of the current thermoelectric materials achieves sufficient thermoelectric efficiency at cryogenic temperature range. Space science applications, cryocooling, and microelectronic superconducting quantum interference devices could be improved by a reliable effective solid state cooling.

Very recently, simply binary compound FeSb$_2$ with an orthorhombic structure has been characterized as an example of strongly correlated non-cubic Kondo insulator with 3d ions \cite{kondo1,kondo2} It was found to exhibit colossal value of Seebeck coefficient at 10 K and a record high thermoelectric power factor $S^2/\rho\sim 2300 \mu WK^{-2}cm^{-1}$ was observed. \cite{fesb1,fesb2,rongwei} This might imply FeSb$_2$-based materials with narrow energy gaps and correlated bands could be good thermoelectrics at cryogenic temperatures. However, the figure of merit (ZT) is low due to the very large thermal conductivity values. On the other hand it was shown that properties of Kondo insulators are sensitive to chemical substitution \cite{doping1, doping2, doping3}. Here, we carefully studied the influence of Te-doping on Sb-site on thermal  and thermoelectric properties.
Only $5\%$ Te doping induces nearly two-order decrease in thermal conductivity and the phonon mean free path, which may be attributed to the substitution disorder. The Te doping also results in transition from an insulator to a metal. Consequently a significantly enhanced thermoelectric figure of merit $(ZT\sim0.05)$ when compared with $ZT<0.005$ in undoped FeSb$_2$ is achieved in Fe(Sb$_{0.9}$Te$_{0.1}$)$_2$.

Fe(Sb$_{1-x}$Te$_x$)$_2$ ($x=0, 0.005, 0.01, 0.05$) single crystals were grown from excess Sb flux, as described previously \cite{fesbte}. The thermal transport properties including thermal conductivity, Seebeck coefficient, as well as electrical resistivity were measured in Quantum Design PPMS-9 from 2K to 300K using one-heater-two-thermometer method. The direction of heat and electric current transports was along the c-axis of the crystals oriented using a Laue camera. The heat capacity of all samples, which used to derive the phonon mean free path, were measured also using Quantum Design PPMS-9 in the same temperature range.

\begin{figure}
\includegraphics[scale=1.2]{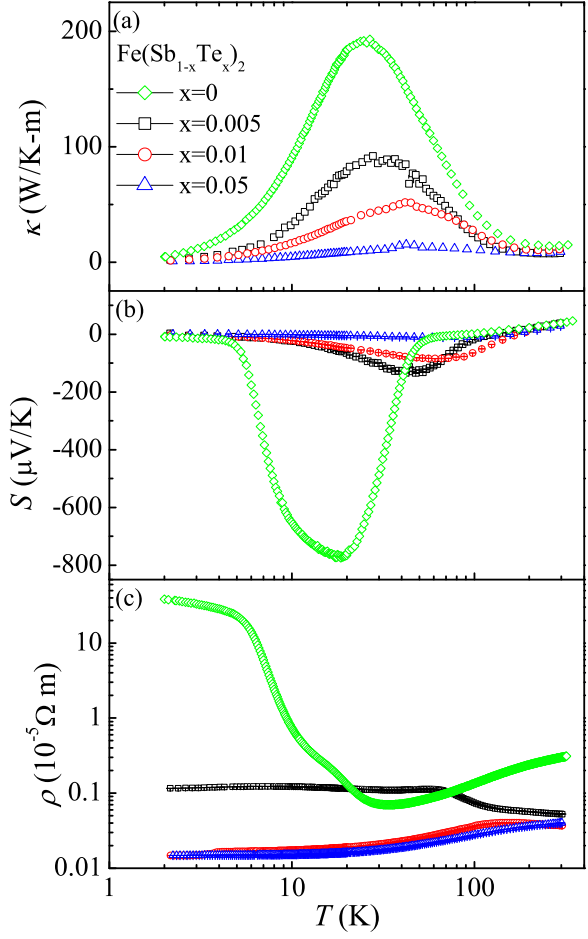}%
\caption{The temperature dependence of thermal conductivity $\kappa$ (upper figure), Seebeck coefficient $S$ (middle figure) and resistivity $\rho$ (bottom figure) of Fe(Sb$_{1-x}$Te$_x$)$_2$ samples with $x=0, 0.001, 0.01, and 0.05$, respectively.\label{}}%
\end{figure}

Figure 1 shows the temperature dependence of the thermal conductivity, Seebeck coefficient and electrical resistivity of samples with different Te doping level. The thermal conductivity $\kappa(T)$ achieves its maximum between 12 and 20K for pure FeSb$_2$, peaking at  ~250 W/K-m. The Seebeck coefficient changes sign from positive to negative around 120K, indicating presence of two carrier types. The absolute value of $S$ increases rapidly below 40K, and reaches its nearly constant maximum value $|S(T)|_{max}\sim800mV/K$ in the temperature interval $20K\sim10K$. This is similar to previous report \cite{fesb1,rongwei}. Due to large thermal conductivity, its $ZT$ ($ZT<0.005$ at 10K) is very low, which seriously limits the realistic applications. A slight substitution of Te at Sb sites ($0.5\%$) significantly suppresses the thermal conductivity by half, i.e., $\kappa_{max}<90 W/K-m$. Further doping with Te induces more significantly decrease in $\kappa$, and only $5\%$ Te doping induces nearly two-order decrease in thermal conductivity. The substitution also induces the significant suppression of the resistivity and a change to metallic ground state. The metallic temperature region increases with Te doping.

\begin{figure}
\includegraphics[scale=1.0]{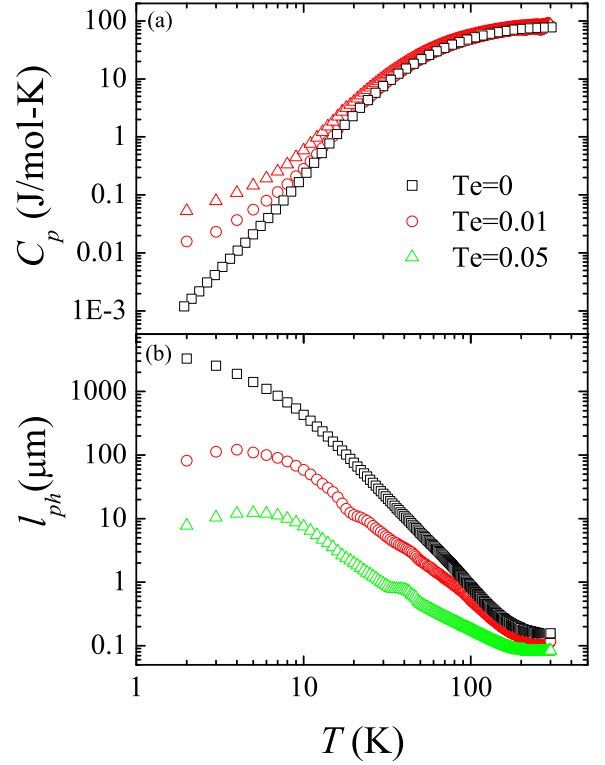}%
\caption{The specific heat (a) and derived phonon mean free path $l_{ph}$ (b) as a function of temperature for samples with $x=0,0.01,0.05$, respectively.\label{}}%
\end{figure}

\begin{figure}
\includegraphics[scale=1.0]{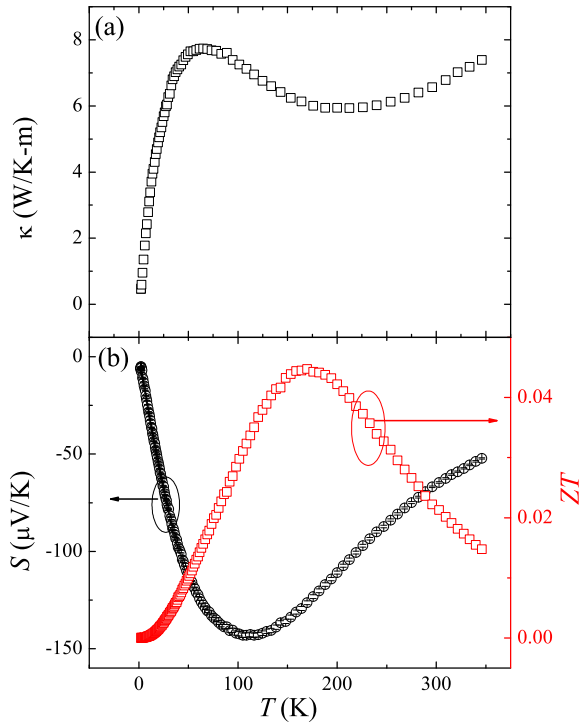}%
\caption{The thermal conductivity (a), Seebeck coefficient and thermoelectric figure of merit (b) of Fe(Sb$_{0.9}$Te$_{0.1}$)$_2$ as a function of temperature.\label{}}%
\end{figure}

In order to clarify the thermal conductivity suppression, we have extracted the phonon mean free path. The carrier contribution to thermal conductivity $\kappa_c$  calculated from the electrical resistivity using Wiedemann-Franz law $\frac{\kappa}{\tau T}=L_0=2.44\times10^{-8}W\Omega K^{-2}$ with $\tau=1/\rho$ shown in Fig. 1(c), is rather small and repesents less then $1\%$ of the total thermal conductivity. Therefore $\kappa$ is dominated by the lattice thermal conductivity below room temperature in all investigated crystals especially in the region of large $S$. We estimated the lattice thermal conductivity $\kappa_L$  by subtracting $\kappa_c$ from the total thermal conductivity. By using the kinetic formula $\kappa_L=\frac{1}{3}C_L\nu l_{phonon}$ where $C_L$ is lattice specific heat (Fig.2(a)) and $\nu$ is the sound velocity,we calculated the phonon mean free path, the $l_{phonon}$ for three samples with Te$=0,0.01,0.05$  (Fig.2(b)). The phonon mean free path is suppressed by nearly two orders of magnitude by Te doping. The phonon mean free path is very sensitive to the grain boundary and disorder effect that could enhance the scattering process and suppress the phonon mean free path. The suppression of $l_{phonon}$ in our system most likely originates from substitutional disorder in doped crystals in the absence of grain boundaries.

The suppression of thermal conductivity and electrical resistivity by Te doping in FeSb$_2$ is expected to enhance the thermoelectric figure of merit. Unfortunately, the Seebeck coefficient is suppressed simultaneously with thermal conductivity by Te doping, as shown in Fig.1(b). The peak of $S(T)$ is also shifts to higher temperature. But it can be expected that for some doping range there is an optimal value of Seebeck coefficient, thermal conductivity and ZT.

Fig.3 (a) shows thermal conductivity of Fe(Sb$_{0.9}$Te$_{0.1}$)$_{0.2}$. It can be concluded that the thermal conductivity is further suppressed. The peak value of $\kappa$ is only $\sim 8 W/K-m$. A maximum of Seebeck coefficient $\sim 150 \mu V/K$ at $\sim 100K$ is observed, as shown in Fig.3 (b). Correspondingly, the maximum of $ZT$ achieves 0.04 at about 100K, which is nearly one-order of magnitude larger than $ZT\sim 0.005$ at undoped FeSb$_2$.

In summary, we studied the influence on the thermal conductivity and thermoelectric effect by Te doping at Sb site in FeSb$_2$ single crystal by the self-flux method.
Only $5\%$ Te doping induces nearly two-order decrease on thermal conductivity and the phonon mean free path. This may be attributed to substitutional disorder.
The Te doping also results in transition from a semiconductor to a metal. Consequently a significantly enhanced thermoelectric figure of merit $(ZT\sim0.05)$ is achieved in Fe(Sb$_{0.9}$Te$_{0.1}$)$_{0.2}$ at $\sim~100K$. Even though this value
is too small to for applications, our study pointed to a direction for the thermal conductivity suppression in FeSb$_2$.

\textit{Note added.} We became aware that a preprint was posted on the arXiv.org with similar conclusion on the same day of our submission. \cite{sun}.

This work was carried out at Brookhaven National Laboratory. Work at Brookhaven is supported by the U.S. DOE under contract No. DE-AC02-98CH10886.


%

%




\begin{thebibliography}{99}
\bibitem{TE1}
G.~D.~Mahan, Solid State Phys., {\bf 51}, 81 (1997).
\bibitem{TE2}
G.~J.~Synder and E.~S.~Toberer, Nature Mater. {\bf 7}, 105 (2008).
\bibitem{phonon1}
B. Paudel, et. al., Science {\bf 320} 634 (2008).
\bibitem{phonon2}
W.~Kim, et. al., Phys. Rev. Lett. {\bf 96}, 045901 (2006).
\bibitem{phonon3}
B.~C.~Sales, et. al. Phys. Rev. B {\bf 56}, 15081 (1997).
\bibitem{FeSi}
A.~Sakai, F.~Ishii, Y.~Onose, Y.~Tomioka, S.~Yotsuhashi, H.~Adachi, N.~Nagaosa, and Y.~Tokura, J. Phys. Soc. Jpn. {\bf 76}, 093601 (2007).
\bibitem{NaCoO}
I. Terasaki, Y. Sasago, and K. Uchinokura, Phys. Rev. B {\bf 56}, R12685 (1997).
\bibitem{correlated1}
J.~M.~Tomczak, K.~Haule, T.~Miyake, A.~Georges, and G.~Kotliar, Phys. Rev. B {\bf 82}, 085104 (2010).
\bibitem{correlated2}
G.~Palsson and G.~Kotliar, Phys. Rev. Lett. {\bf 80}, 4775 (1998).
\bibitem{correlated3}
R.~Arita, K.~Kuroki, K.~Held, A.~V.~Lukoyanov, S.~Skornyakov, and V.~I.~Anisimov, Phys. Rev. B {\bf 78}, 115121 (2008).
\bibitem{kondo1}
C.~Petrovic, J.~W.~Kim, S.~L.Bud'ko, A.~I.~Goldman, P.~C.~Canfield, W.~Choe, and G.~J.Miller, Phys. Rev. B {\bf 67}, 155205 (2003).
\bibitem{kondo2}
C.~Petrovic, Y.~Lee, T.~Vogt, N.~D.~Lazarov, S.~L.~Bud'ko, and P.~C.~Canfield, Phys. Rev. B {\bf 72}, 045103 (2005).
\bibitem{fesb1}
P.~Sun, N.~Oeschler, S.~Johnsen, B.~B.~Iversen, and F.~Steglich, Phys. Rev. B {\bf 79}, 153308 (2009).
\bibitem{fesb2}
A.~Bentien, S.~Johnsen, G.~K.~H. Madsen, B.~B.~Iversen, and F.~Steglich, Eur. Phys. Lett. {\bf 80}, 17008 (2007).
\bibitem{fesbte}
R.~Hu, V.~F.~Mitrovi\'{c}, and C.~Petrovic, Phys. Rev. B {\bf 79}, 064510 (2009).
\bibitem{rongwei}
Q.~Jie, R.~Hu, E.~S.~Bozin, C.~Petrovic, and Q.~Li, Unpublished.
\bibitem{doping1}
R.~Hu, V.~F.~Mitrovic, and C.~Petrovic, Phys. Rev. B {\bf 79}, 064501 (2009).
\bibitem{doping2}
R.~Hu, V.~F.~Mitrovic, and C.~Petrovic, Phys. Rev. B {\bf 76}, 115105 (2009).
\bibitem{doping3}
N.~Manyala, Y.~Sidis, J.~F.~Ditusa, G.~Aeppli, D.~P.~Young, and Z.~Fisk, Nature Mater. {\bf 3}, 255 (2004).
\bibitem{sun}
P.~Sun, M.~Sondergaard, Y.~Sun, S.~Johnsen, B.~B.~Iversen, and F.~Steglich, arXiv:1102.0171 (2011).
\end{thebibliography}

\end{document}